**Title:** Phase imaging in scanning transmission electron microscopy using bright-field balanced divergency method

**Authors:** Binbin Wang[a,b1] and David W. McComb [a,b,*]

**Affiliations:**

[a] Center for Electron Microscopy and Analysis, the Ohio State University, Columbus, Ohio 43212, US; and [b] Material Science and Engineering, the Ohio State University, Columbus, Ohio 43212, US

*David W. McComb

**Email:** mccomb.29@osu.edu

[1] Present address: Intel Corporation, Hillsboro, Oregon, 97124, US

**Abstract:** We introduce a phase imaging mechanism for scanning transmission electron microscopy that exploits the complementary intensity changes of transmitted disks at different scattering angles. For scanning transmission electron microscopy, this method provides a straightforward, dose-efficient, and noise-robust phase imaging, from atomic resolution to intermediate length scales, as a function of scattering angles and probe defocus. At atomic resolution, we demonstrate that the phase imaging using the method can detect both light and heavy atomic columns. Furthermore, we experimentally apply the method to the imaging of nanoscale magnetic phases in FeGe samples. Compared with conventional methods, phase retrieval using the new method has higher effective spatial resolution and robustness to non-phase background contrast. Our method complements traditional phase imaging modalities in electron microscopy and has the potential to be extended to other scanning transmission techniques and to characterize many emerging material systems.

## 1. Introduction

Phase contrast imaging is a key research subject in many fields, including optical metrology, electron microscopy, X-ray optics, condensed matter physics and biomedical imaging. In the case of transmission electron microscopy (TEM), most samples of interest can be viewed as phase objects, where the complex amplitude distribution in the object plane is convolved into the wave function of the transmitted electron beam [1]. However, only the amplitude variation, *i.e.* intensity, rather than phase variation in the transmitted electron beam is directly detected. In TEM mode, to recover phase information in the transmitted electron beam requires approaches based on interferometry or electron beam propagation [2], such as phase plate methods [3,4], off-axis electron holography [5] and transport of intensity equation (TIE) methods [6,7]. Most of these techniques require complicated optical configurations and alignments, and/or image registration, which are sensitive to external disturbances (*e.g.* optical changes, vibration, camera noises), limiting imaging resolution and quality. In scanning TEM (STEM), phase imaging can instead be retrieved by the center of mass (CoM) shift of the transmitted electron beam, *i.e.* differential phase contrast (DPC) [8,9], or by electron ptychography [10–12]. However, the phase retrieved using the DPC method is often sensitive to probe shape effects, optical changes, specimen bending/thickness changes, or sharp potential changes in the object. While electron ptychography shows exquisite resolution it requires elaborate optimization of experimental and reconstruction parameters which can be limiting for less experienced researchers.

More efficient phase recovery imaging methods are a subject of intense interest as a path to further improve dose efficiency, signal-to-noise ratio (SNR), fidelity and simplicity of phase reconstruction. The recent advent of the fast direct electron detectors has stimulated intense research activity in 4D-STEM methods where spatially resolved diffraction patterns are recorded in scanning mode [13,14]. Since the diffraction patterns contain all transmitted electrons, unconventional multipurpose reconstructions can be built from these data in post-processing to extract signals of interest, such as virtual imaging, strain mapping, symmetry STEM [15], CoM-DPC [16], matched illumination and detector interferometry STEM [17] and mixed-state electron ptychography [18].

In this contribution, we propose a new approach entitled Bright-field Balanced Divergency (BBD) imaging. This is a straightforward phase contrast retrieval for defocused STEM based on the symmetry of the phase contrast transfer function (PCTF) of virtual bright-field (BF) and its complementary annular bright-field (ABF) images. We will show that the BBD method is non-iterative, does not require registration of complimentary images and is noise-robust. In the weak absorption approximation, both overfocused and underfocused BF/ABF STEM images can be virtually reconstructed at same time from a single defocused STEM dataset, due to the complementary intensity changes of the transmitted disk at different scattering angles. Reliable phase contrast can be reconstructed from these images, providing a way to study the structure or the magnetic/electrical phase in the sample, with many advantages over traditional methods.

## 2. Theory and methods

### 2.1. Experimental and analysis

All experimental results presented in this paper were performed on a FEI 300 kV image-corrected Titan G2 60−300 S/TEM, equipped with an EMPAD [19]. FeGe samples for characterization were prepared from single crystals using a FEI Helios NanoLab 600 DualBeam focused ion beam (FIB). The sample temperature was controlled using a liquid nitrogen cryo-holder (Gatan 636), while the magnetic field was applied in Lorentz mode by adjusting the objective lens current of the microscope. TEM images were recorded using a Gatan UltraScan CCD, in which we controlled the electron dose by defocusing the monochromator. 4D-STEM datasets of FeGe samples were recorded using the EMPAD with dwell time of 1 ms while the convergence angle and dose were adjusted as necessary by changing the aperture and defocusing the monochromator, respectively.

Postprocessing for virtual detectors and CoM-DPC was done using custom scripts in python based on pixSTEM [20] and py4DSTEM [21]. Note that our previous work showed that spatial frequency filters can help filter out special frequencies in CoM-DPC images by choosing appropriate convergence and collection angles [22]. To address the optimal spatial resolution in CoM-DPC, here we use (3/4-1)$\alpha$ for image reconstruction. Phase and field imaging using the BBD method was done using a custom script in MATLAB based on the open source package of TIE's FFT solution given in ref. [6].

**2.2. CTF calculations and STEM Simulations**

A model detailing the PCTF calculation in this paper is given in Appendix A. MuSTEM was used to simulate images of $MoS_2$ mono/multilayers [23]. The parameters are as follows: a periodic orthogonal 3×3×1 $MoS_2$ supercell with vacancies is created by deleting one S atom from the top layer; the monolayer $MoS_2$ is sliced into 3 layers for modeling; the Debye-Waller factors at room temperature is from reference [24]; the accelerating voltage is 80 kV; probe convergence semi-angle is 30 mrad; for simplicity, defocus is changed as needed without considering residual spherical aberration; built-in virtual detectors are defined to create BF, ABF and ADF images of different collection angle ranges. For $MoS_2$ multilayer, we repeat the $MoS_2$ monolayers with single S vacancy by setting the layer distance to be 6.2 Å.

**3. Results**

**3.1. Calculations.** Figure 1a shows a schematic of a STEM defocus experiment. Instead of focusing the electron probe on the sample, we set the focal plane a distance away from the sample ($\Delta f \neq 0$). As illustrated in Figure 1a, the scattering intensity distribution in the diffraction plane takes the form of a convergent beam diffraction pattern (CBED), from which virtual BF or ABF images can be obtained by integrating electrons over a specific range of scattering angles. The intensity distribution of the CBED patterns varies with probe position, defocus value, accelerating voltage, and convergence angle. To illustrate the BBD method, STEM simulations were performed on a test case of molybdenum disulfide ($MoS_2$) with a single S vacancy site. $MoS_2$, a *Van der Waals* material with a hexagonal plane of Mo atoms between two hexagonal planes of S atoms, has recently been experimentally and computationally explored for different measurements in STEM mode [25,26]. Figure 1b shows the array of images reconstructed from the calculated CBED patterns of a $MoS_2$ monolayer with different defocus ($\Delta f$) and collection angle ranges ($\beta$) for a convergence angle ($\alpha$) of 30 mrad. The aligned images illustrate how contrast varies with $\Delta f$ and $\beta$. For example, when $\Delta f$ is 0, the reconstructed images at all $\beta$ values show negative contrast (dark spots) at atomic positions due to electron scattering. However, changing $\Delta f$ introduces contrast changes including invisibility and contrast inversion. These changes are sensitive to $\beta$ and are approximate symmetric about $\Delta f$. To understand the contrast characteristics at different $\Delta f$ and $\beta$, a multi-slice calculation of PCTF, $\mathcal{L}_\beta(\Delta f, \Delta z, M)$, for the thickness of Δz with M different slices was done with the results shown in Figure 1c (please see Appendix A1).

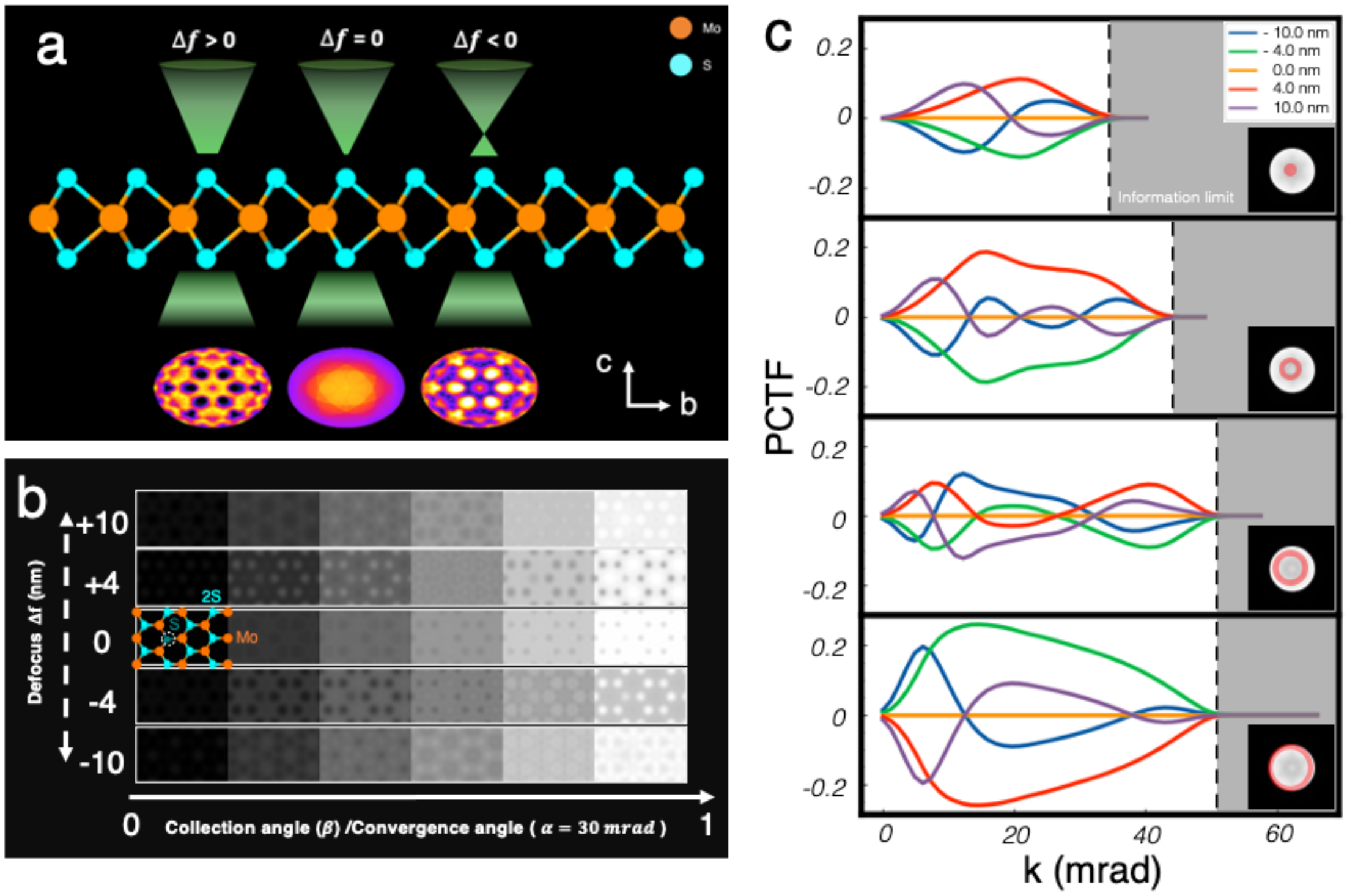

**Figure 1.** Schematic and example of phase contrast evolution in BF/ABF images of monolayer $MoS_2$ with different defocus and collection angle ranges. (a) Modeling setup for different defocusing and examples of varied patterns in transmitted disks associated with defocusing. (b) Simulated BF/ABF STEM images with different defocus ($\Delta f = 0, \pm 4, \pm 10$ nm) and collection angle β ranges at 80 kV with a step of 5 mrad and convergence angle α of 30 mrad. Insert is a $MoS_2$ unit cell with a single S defect (dimmer) in the white circle. The unit cell is tilted a few degrees from the c-axis to illustrate overlapping S atoms. (c) Calculated PCTF associated with (b). The insert in (c) provides the selected scattering angle ranges relative to the transmitted disk α: (0-1/4)α, (1/4-1/2)α, (1/2-3/4)α and (3/4-1)α. Vertical dashed lines indicate the information limit.

***3.1.1 Dependence on defocus.*** The PCTF represents how object information appears in the reconstructed image according to the spatial frequency. For example, using the annulus with inner and outer β values of ¾α and 1α, respectively, when $\Delta f$ is -4 nm, the sign of PCTF is robustly positive for all spatial frequencies below ~50 mrad. Therefore, the virtual ABF image of this detector configuration is imaged with positive contrast, which means the atomic positions will appear as bright spots as shown in Figure 1b. However, when $\Delta f$ = -10 nm, the sign of the PCTF is inverted at a specific spatial frequency. In this case the virtual ABF image has no simple correlation with the real atomic structure. The same trend can be observed when Δf is positive, since the PCTF simply reverses when thickness effects are negligible. To avoid this spatial frequency dependent PCTF inversion, $|\Delta f|$ should be smaller than $\lambda^{-1}|\boldsymbol{k}|^{-2}$ under the paraxial approximation ($\lambda^2|\boldsymbol{k}|^2 \ll 1$) (see Appendix A2), where $\lambda$ is the wavelength of the electron and **k** is the coordinate in reciprocal space. Figure 2a shows the calculated $\lambda^2|\boldsymbol{k}|^2$ and the critical defocus, *i.e.* $\lambda^{-1}|\boldsymbol{k}|^{-2}$, with different $\boldsymbol{k}$ for accelerating voltages of 80 kV and 300 kV, respectively. The critical defocus is inversely proportional to $|\boldsymbol{k}|^2$ and $\lambda$: for atomic resolution imaging (~10 mrad), the critical focus is ~20 nm at 300 kV, and for features > 10 nm (100 μrad) the critical focus is ~400 μm. Note that the absorption and scattering effects are not considered in the calculation of PCTF, which results in weaker negative contrast in conventional ABF images when $\Delta f$ is 0.

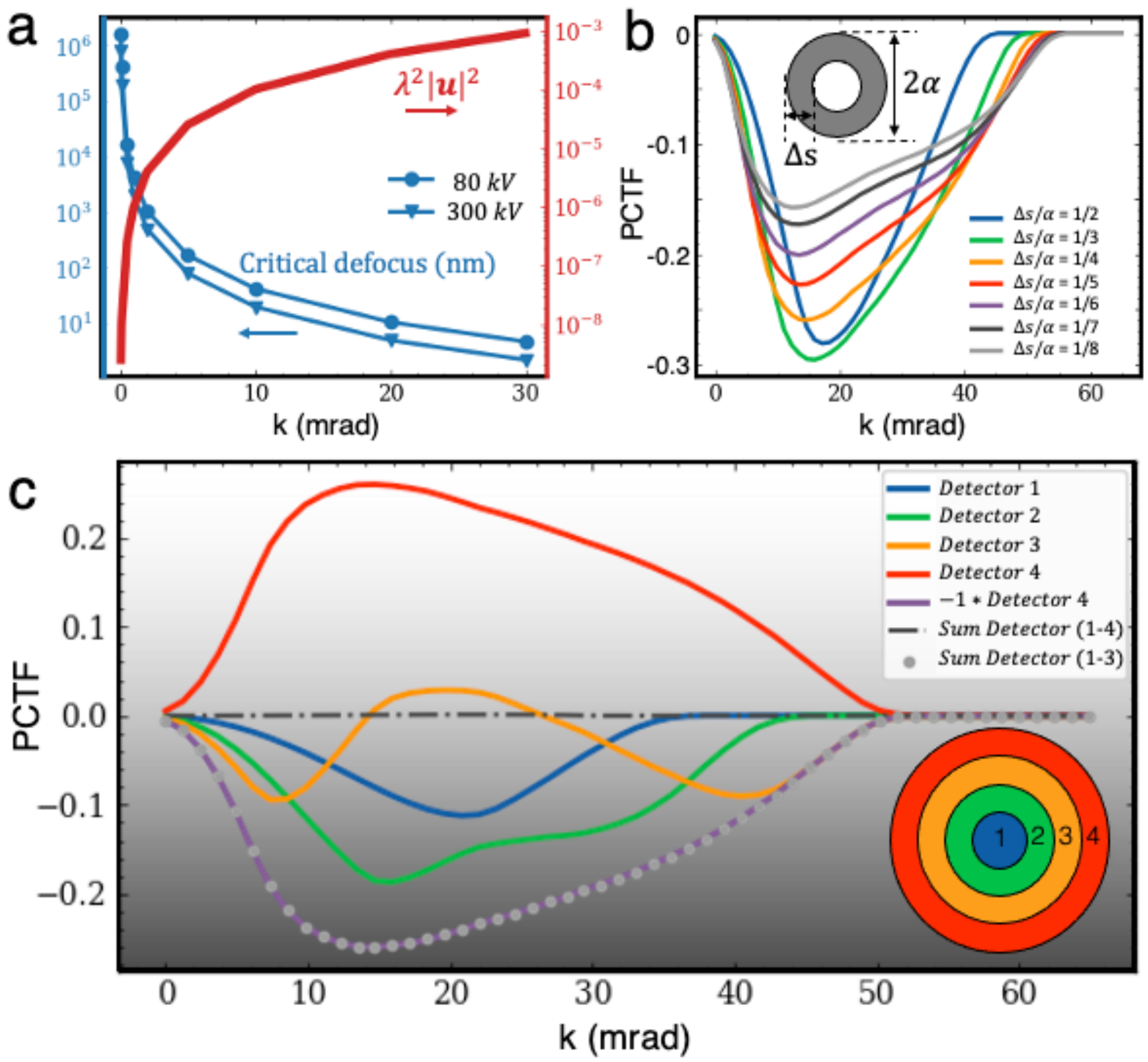

**Figure 2.** Dependence on collection angle ranges. (a) Critical defocus and paraxial approximation with different **k**. (b) PCTF of defocus 4 nm at 80 kV for various virtual detectors. Insert is a schematic of these virtual detectors, which have the same maximum annular outer α, 30 mrad, but different widths Δs. (c) PCTF of defocus -4 nm at 80 kV with 4 virtual segmented detectors and their combinations: (0-1/4)α, (1/4-1/2)α, (1/2-3/4)α, (3/4-1)α, inverted (3/4-1)α, (0-1)α and (0-3/4)α.

***3.1.2. Dependence on collection angle ranges.*** In addition to defocus dependence**,** the PCTF also varies with the collection angle range, as illustrated in Figure 1b and 1c. The PCTF expands to higher spatial frequencies as the inner angle of the virtual detector increases. Consequently, images reconstructed at higher collection angle ranges have higher information limit, *i.e.* higher resolution, than images reconstructed at lower angular ranges. The same trend is also observed in Figure 2b for PCTF with a defocus of 4 nm at 80 kV for various virtual detectors with a fixed annular outer angle $\alpha$, 30 mrad, but different widths $\Delta s$ (the difference between inner and outer collection angles). The information limit is around 45 mrad at $\Delta s/\alpha$ = 1/2 and converges to be 55 mrad as $\Delta s/\alpha \rightarrow 0$. Meanwhile, when $\Delta s/\alpha$ < 1/3, the amplitude of PCTF in Figure 2b is reduced with decreasing $\Delta s$. Changing defocus and spherical aberration will not change the trend, as shown in Appendix B, Figure S1.

Therefore, the $\Delta s/\alpha$ of the optimized virtual disk should be between 1/3-1/4 to achieve high imaging resolution and strong phase contrast at the same time. Using the TIE, the pure phase contrast image $\phi(\boldsymbol{r})$ can be approximately reconstructed from virtual ABF images acquired at $\pm\Delta$f by (please see Appendix A2):

$$\phi(\boldsymbol{r}) = \frac{|I_\beta(\boldsymbol{r},\Delta \mathrm{f}) - I_\beta(\boldsymbol{r},-\Delta \mathrm{f})|}{4\pi\lambda\Delta f|\boldsymbol{k}|^2[I_\beta(\boldsymbol{r},\Delta \mathrm{f}) + I_\beta(\boldsymbol{r},-\Delta \mathrm{f})]} \quad [1]$$

where $I_\beta(\boldsymbol{r},\Delta \mathrm{f})$, $\lambda$ and $\boldsymbol{k}$ are the reconstructed images from detector $D_\beta$ at defocus Δf, electron wavelength, and the coordinates in the reciprocal space, respectively.

***3.1.3. Bright-field balanced divergency.*** The PCTF of a virtual ABF image from $(\alpha - \Delta s, \alpha)$ is inversely equal to the one of its complementary virtual BF image from (0, $\alpha - \Delta s$) in opposite $\Delta f$ (see Appendix A3 for more details). As shown in Figure 2c, the inverted PCTF of the virtual ABF image from (3/4-1)$\alpha$ is overlapped with that of the virtual BF image from (0-3/4)$\alpha$, while the PCTF from (0-1)$\alpha$ does not transfer phase contrast. We refer this result as "bright-field balanced divergency (BBD)", which is based on the total intensity of the transmitted disk being approximately constant at each scan position, which depends on the absorption variable η(k) but is approximately independent of Δf under the paraxial approximation. Therefore, the PCTF amplitudes of both virtual detectors in Figure 2b and its complementary detectors will decrease to 0 as $\Delta s \to 0$. Mathematically, we can define two virtual detectors in the transmitted disk circle:

$$\begin{cases} D_{\beta 1}(\boldsymbol{k}) = \begin{cases} 1, & \boldsymbol{k} < \beta 1 \\ 0, & \beta 1 < \boldsymbol{k} < \alpha \end{cases} \\ D_{\beta 2}(\boldsymbol{k}) = \begin{cases} 0, & \boldsymbol{k} < \beta 1 \\ 1, & \beta 1 < \boldsymbol{k} < \alpha \end{cases} \end{cases}.$$

When $\Delta f$ or **k** is small, *i.e.* ignoring higher-order expansion terms in the paraxial approximation ($\lambda^2|\boldsymbol{k}|^2 \ll 1$), the expression for the reconstructed image from these two detectors can be simplified to

$$\begin{cases} I_{\beta 1}(\boldsymbol{r},\Delta f) = I_{0,\beta_1}(\boldsymbol{k}) - 2\left\{\int_0^{\beta_1} e^{2\pi i \boldsymbol{k}\cdot\boldsymbol{r}} \eta(\boldsymbol{k}) d\boldsymbol{k}\right\} - 2\left\{\int_0^{\beta_1} \sin(\pi\Delta f\lambda|\boldsymbol{k}|^2) e^{2\pi i \boldsymbol{k}\cdot\boldsymbol{r}} \phi(\boldsymbol{k}) d\boldsymbol{k}\right\} \\ I_{\beta 2}(\boldsymbol{r},\Delta f) = I_{0,\beta_2}(\boldsymbol{k}) - 2\left\{\int_{\beta_1}^{\alpha} e^{2\pi i \boldsymbol{k}\cdot\boldsymbol{r}} \eta(\boldsymbol{k}) d\boldsymbol{k}\right\} - 2\left\{\int_{\beta_1}^{\alpha} \sin(\pi\Delta f\lambda|\boldsymbol{k}|^2) e^{2\pi i \boldsymbol{k}\cdot\boldsymbol{r}} \phi(\boldsymbol{k}) d\boldsymbol{k}\right\} \end{cases}.$$

Since they are "bright-field balanced", we have

$$\left\{\int_0^{\beta_1} \sin(\pi\Delta f\lambda|\boldsymbol{k}|^2) e^{2\pi i \boldsymbol{k}\cdot\boldsymbol{r}} \phi(\boldsymbol{k}) d\boldsymbol{k}\right\} + \left\{\int_{\beta 1}^{\alpha} \sin(\pi\Delta f\lambda|\boldsymbol{k}|^2) e^{2\pi i \boldsymbol{k}\cdot\boldsymbol{r}} \phi(\boldsymbol{k}) d\boldsymbol{k}\right\} = 0$$

$$\Rightarrow \left\{\int_0^{\beta_1} \sin(\pi\Delta f\lambda|\boldsymbol{k}|^2) e^{2\pi i \boldsymbol{k}\cdot\boldsymbol{r}} \phi(\boldsymbol{k}) d\boldsymbol{k}\right\} = \left\{\int_{\beta 1}^{\alpha} \sin(\pi(-\Delta f)\lambda|\boldsymbol{k}|^2) e^{2\pi i \boldsymbol{k}\cdot\boldsymbol{r}} \phi(\boldsymbol{k}) d\boldsymbol{k}\right\}.$$

Importantly, we can replace the $I_\beta(\boldsymbol{k}, -\Delta \mathrm{f})$ in Eq.1 with the image reconstructed from the complementary detector of $D_\beta$ at $\Delta f$. Then we can get

$$\phi(\boldsymbol{r}) \approx \frac{|\gamma * I_{\beta 1}(\boldsymbol{r},, \Delta f) - I_{\beta 2}(\boldsymbol{r},, \Delta f)|}{2 * (1+\gamma)[I_{\beta 1}(\boldsymbol{r}, \Delta f) + I_{\beta 2}(\boldsymbol{r}, \Delta f)] \sin(\pi \Delta f \lambda |\boldsymbol{k}|^2)},$$

where $\gamma = \frac{I_{\beta 2}(\boldsymbol{r}, \Delta f=0)}{I_{\beta 1}(\boldsymbol{r}, \Delta f=0)} \approx \frac{I_{\beta 2}(\boldsymbol{r}, \Delta f)}{I_{\beta 1}(\boldsymbol{r}, \Delta f)}$ in weak phase approximation, which is to balance the dose difference in virtual images.

This means, phase-contrast images can be reconstructed using the BBD method from the same datasets with different virtual detectors but the same defocus, compared to traditional methods that require acquiring a series of images with different defocus.

**3.2. Phase imaging of $MoS_2$.** To demonstrate phase reconstructions using the BBD method, we used BF and ABF images of $MoS_2$ mono/multilayers simulated using the parameters described in Methods. Projected potential (Figure 3a) and HAADF (Figure 3b) images were generated for reference. The phase and electrostatic field imaging in Figure 3C-E were reconstructed from ABF and BF images with |Δf| = 4 nm. As expected, phase imaging using the BBD method in monolayer $MoS_2$ with Δf exhibited atomic-scale contrast and sensitivity to both light ($Z_S$ = 16) and heavy ($Z_{Mo}$ = 42) atoms. For comparison, the single S atom in the HAADF image is barely visible. The intensity ratio of a single S and Mo atom, $(I_S/I_{Mo})$, calculated from the line profile in Figure 3f is 0.22 for the HAADF image, *i.e.* $I_S/I_{Mo} \approx (Z_S/Z_{Mo})^{1.6}$, and 0.67 for the image reconstructed via BBD method, *i.e.* $I_S/I_{Mo} \approx (Z_S/Z_{Mo})^{0.4}$. With increase of thickness, however, the local maxima at atomic sites in the phase image obtained with BBD method (Figure 3d) become sharper compared to monolayer $MoS_2$ (Figure 3c). This is particularly the case for the peaks corresponding to Mo columns with narrow full width half maxim in Figure 3f, although an "false" contrast in the center of Mo-S hexagons appears in association with channeling effect. Even that, the phase imaging using BBD method have better contrast when compared to the conventional focused series method (Figure 3e). Thickness effects should be considered to understand the improvement in spatial resolution of multilayer samples. As shown in the inserts in Figure 3c-e, the "effective focus" distance between the focus plane and the upper/lower surfaces of the object can vary with different focusing conditions, which results in different evolution of PCTF with thickness at $\pm$Δf, as shown in Figure 3g and Appendix B, Figure S2. Notably, the bright-field balanced divergency between the virtual detector (0-3/4)$\alpha$ and (3/4-1)$\alpha$ is not sensitive to sample thickness and spherical aberration, showing advantages over conventional methods using defocused series. To be clear, atomic resolution imaging still requires aberration correction, as it can introduce an inversion of the CTF with increasing thickness (Appendix B, Figure S2).

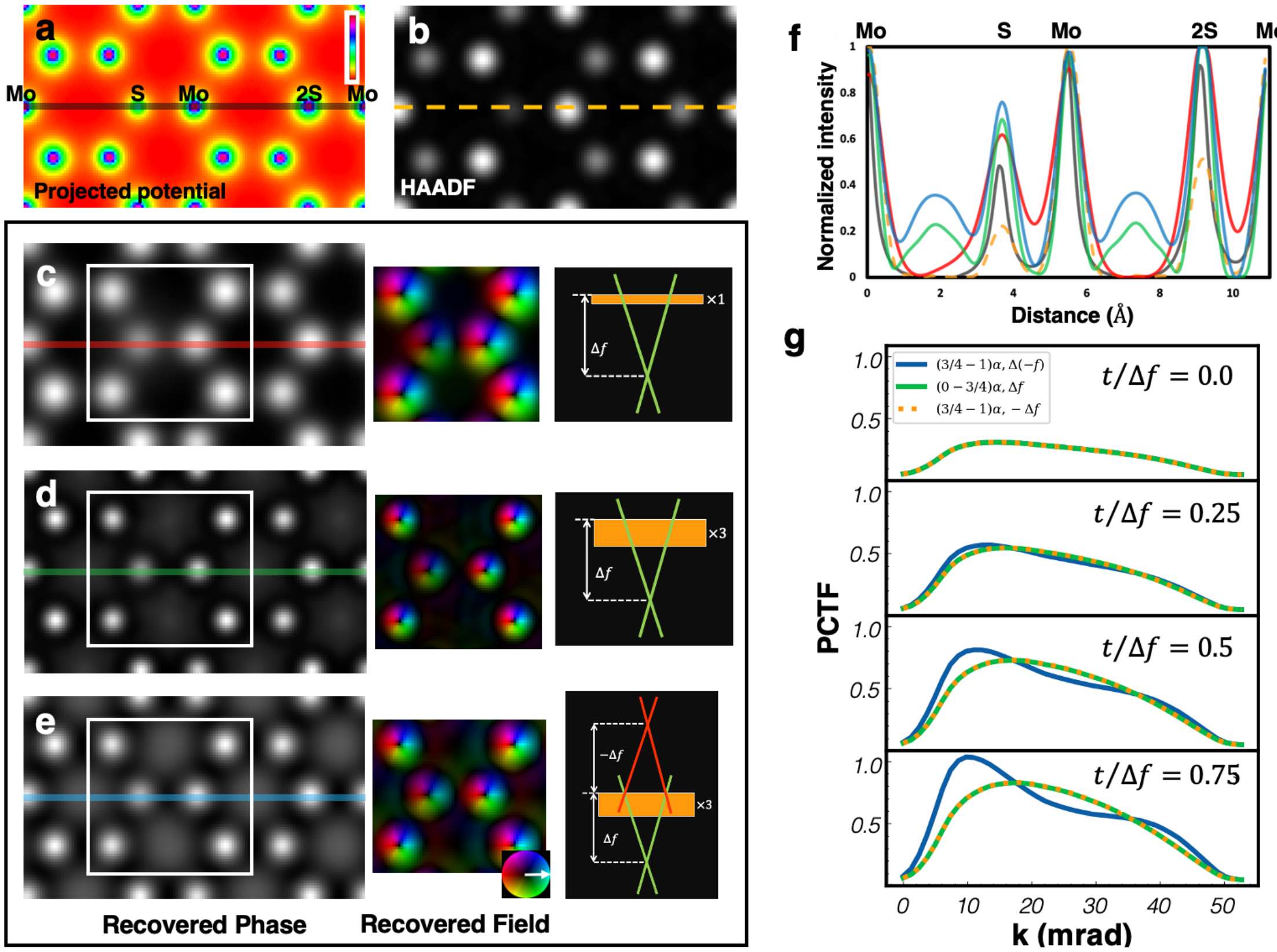

**Figure 3.** Phase imaging of $MoS_2$ monolayer and multilayers at 80 kV with a convergence angle α of 30 mrad. (a) Projected potential image of monolayer $MoS_2$ with single S defect, convolved with Gaussian source-size of 80 pm. (b) HAADF image reconstructed from a monolayer $MoS_2$ using a virtual detector of 50-200 mrad and zero defocus. (c) Phase and projected electrostatic field images of monolayer $MoS_2$ using BBD method with Δf of 4 nm. (d) Phase and projected electrostatic field image of multilayer $MoS_2$ (x3 monolayer) using BBD method with Δf of 4 nm. Virtual detectors using for BBD method are (0-3/4)α and (3/4-1)α. (e) Phase and projected electrostatic field image of multilayer $MoS_2$ (x3 monolayer) using conventional method with $\pm\Delta f$ of 4 nm with virtual detector (3/4-1)α. Inserts in (c-e) are the schematic for modeling setups. (f) Comparison of line profiles along the Mo-S-Mo-2S-Mo peak in (a-e). Line profiles are normalized to 0-1 which correspond to the lines inserted in (a-e). (g) Comparison of PCTF of the virtual detector (3/4-1)α with defocus of $\pm\Delta f$ and the virtual detector (0-3/4)α with defocus $+\Delta f$ (|Δf| is 4 nm) for different sample thicknesses t. Note the shift of PCTF to lower spatial frequencies, associated with the "false" contrast in (d) and (e). Color wheel in (c-e) indicates the field direction.

### 3.3. Phase imaging from experimental Lorentz 4D-STEM data

Using the BBD method we examined projected magnetic phase (PMP) imaging from experimental Lorentz 4D-STEM data collected from FeGe lamellae with different thicknesses (Figure 4). FeGe is a helical magnet where magnetic skyrmions, particle-like topological protected magnetic structures, can exhibit a hexagonal lattice in response to external temperature and magnetic fields [27]. Scanned CBED data were collected using a pixelated electron microscope pixel array detector (EMPAD) at 300 kV [14], see Methods. The sample temperature was controlled using a liquid nitrogen cryo-holder (Gatan 636), while the magnetic field was varied by gradually adjusting the objective lens current of the microscope. The convergence angle was ~300 μrad and the defocus Δf was ~150 μm, which satisfies the paraxial approximation. Conventional BF (0 - 1)$\alpha$ image and CoM-DPC (¾ - 1)$\alpha$ image at zero defocus were reconstructed (Figure 4a and 4b, respectively) for comparison with phase and projected magnetic induction field (PMIF) imaging using BBD method. Figure 4d is virtual BF and ABF STEM images used for the phase reconstruction in Figure 4d, where a mixture of magnetic skyrmion domains and helical phases can be seen. Comparing between Figure 4b and 4d, phase and PMIF imaging using the BBD method exhibit a clear contrast, revealing distinct magnetic textures that are robust to bending artifacts and artificial thickness variations. Although the CoM-DPC image also showed magnetic textures, their fine structures are obscured by the non-magnetic background contrast caused by bending and artificial structures in the samples. For example, according to coincidence site lattice theory, the most stable grain boundaries in hexagonal crystal system is Σ7, and it is found in skyrmion lattices [28]. Using BBD method, we clearly show the detailed size and shape of individual skyrmions at a Σ7 boundary with distorted cores of the five-, seven- and ninefold coordinate structure units (as highlighted in Figure 4d and 4e). Such peculiar structural relaxation of skyrmion domain boundaries might be an essential characteristic of magnetic skyrmions and may play an important role in future device applications.

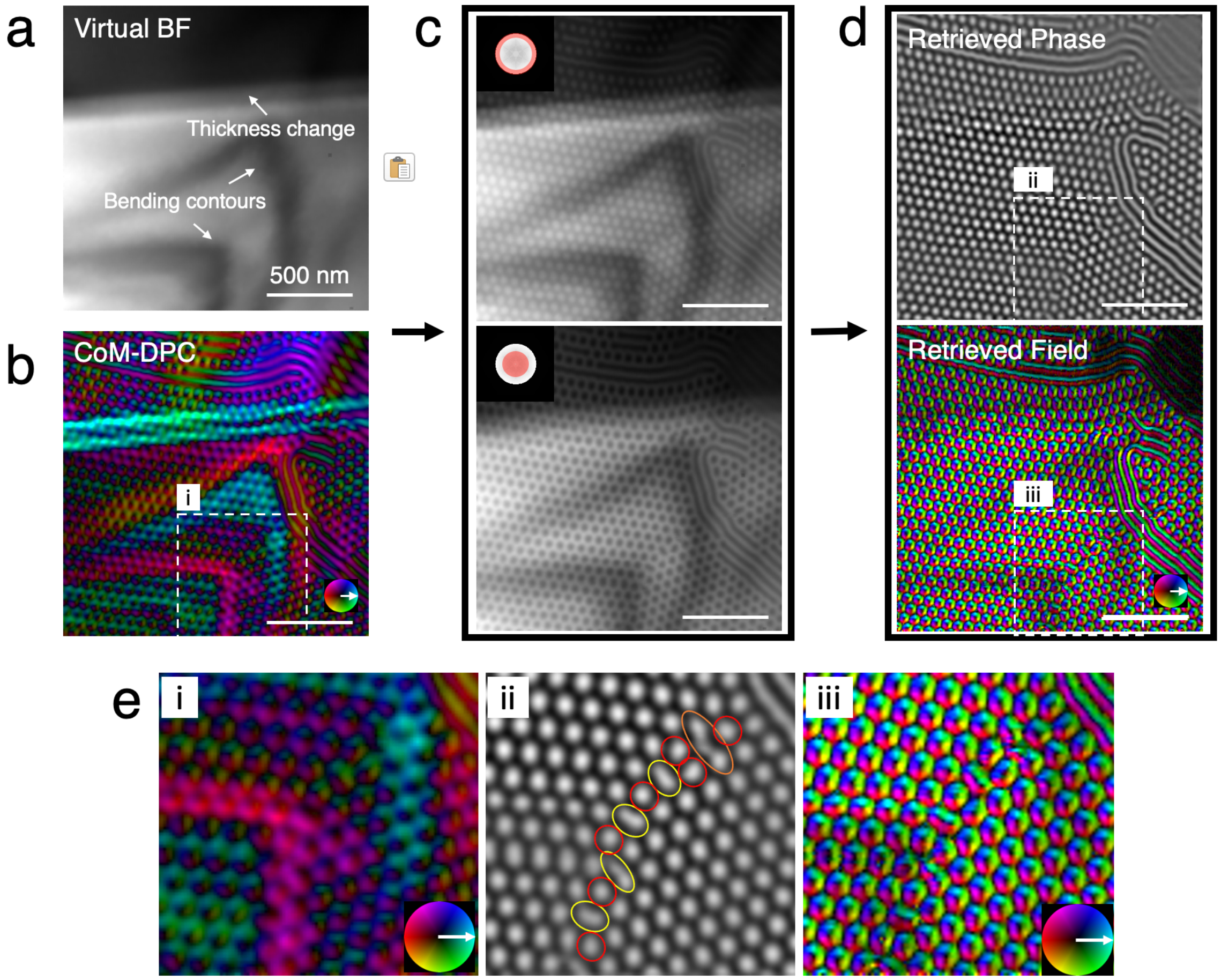

**Figure 4.** PMP and PMIF imaging from Lorentz-4DSTEM data that acquired from FeGe single crystal at 256 K, 90 mT and 300 kV. (a) Virtual BF image reconstructed from angle range (0-1)α with a defocus of 150 μm. (b) The PMIF image reconstructed using CoM-DPC from angle range (3/4-1)α with zero focus. (c) BF and ABF images reconstructed from angle range (0-3/4)α and (3/4-1)α, respectively. (d) Phase and PMIF imaging using BBD method from images in (c). (e) Regions highlighted in (b) and (d) with white boxes, showing the distorted core structures of skyrmions Σ7 domain boundary that cannot be resolved in CoM-DPC. Red, yellow, and orange circles highlight the core skyrmions of five-, seven- and ninefold coordinated structure units, respectively. Color wheels indicate the field direction.

In Figure 5, we compare the performance of reconstructing Lorentz-4DSTEM data using the BBD method with reconstructing Lorentz-TEM images using TIE method. Conventional BF (0-1)$\alpha$ and zero defocus CoM-DPC (¾ -1)$\alpha$ images (Figure 5a and Appendix B, Figure S3, respectively) were reconstructed for comparison. Again, magnetic textures in CoM-DPC image (Appendix B, Figure S3) cannot be well resolved in this sample due to the strong non-magnetic background contrast, especially in engineered channels. Figure 5b is the PMIF imaging from Lorentz-TEM images acquired at Δf=$\pm 150$ μm, while Figure 5c and 5d are PMP and PMIF imaging from Lorentz-STEM data that acquired at Δf =+150 μm. The reconstruction using the BBD method has a better SNR and effective spatial resolution than the conventional reconstruction using TIE method. For example, the helical phase in Figure 5d is faintly visible in Figure 5b, where the magnetic features overlap with labyrinth artifacts caused by camera noises and system misalignments.

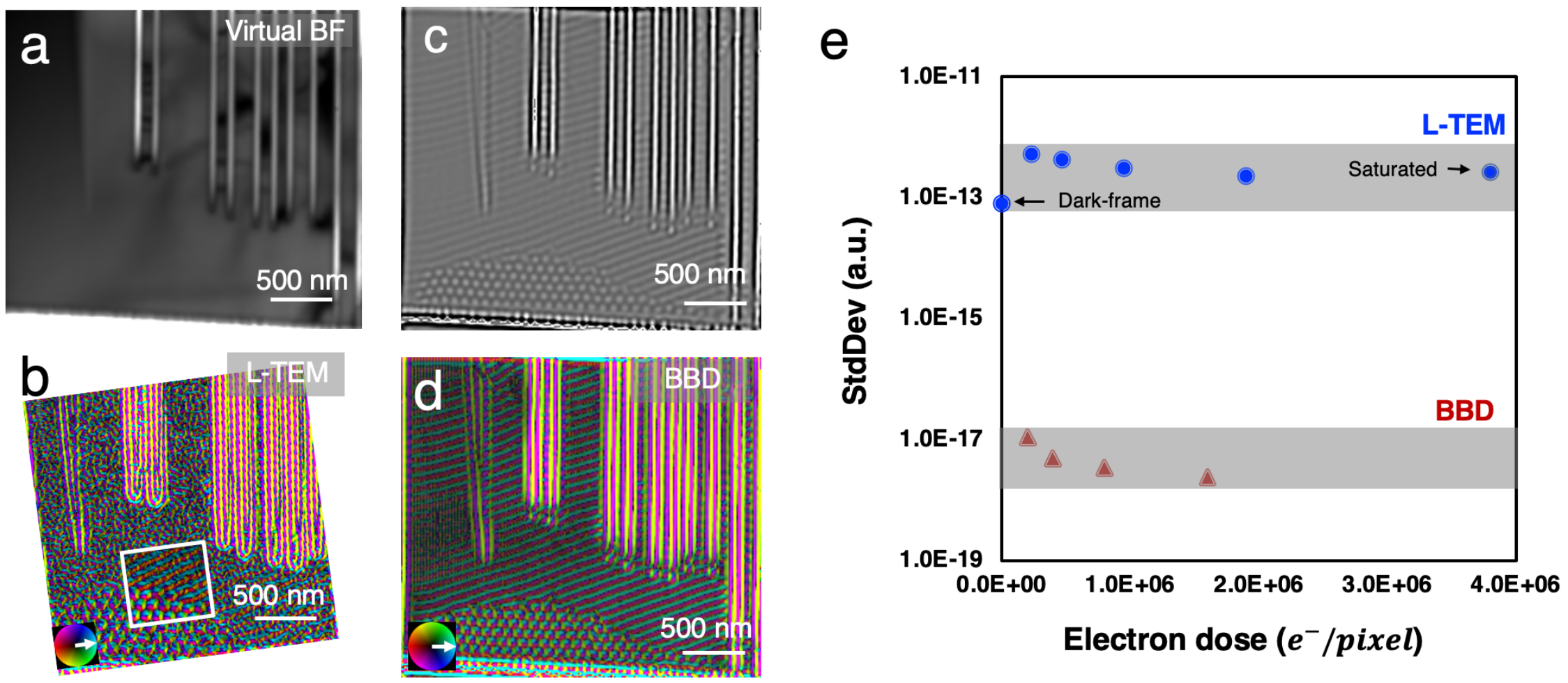

**Figure 5.** Comparison between PMP and PMIF imaging from Lorentz-4DSTEM data using BBD method and Lorentz-TEM images that acquired from geometric confined FeGe at 256 K, 90 mT and 300 kV. (a) Virtual BF image reconstructed from the angle range (0-1)α. (b) Magnetic induction imaging from Lorentz-TEM images without low-pass filter. The white box in (b) is the image with low-pass filter. PMP image (c) and PMIF image (d) from Lorentz-4DSTEM data using BBD method without filtering. Color wheels indicate the field direction. (e) Standard deviation of PMP images with different electron doses from LTEM images and 4D-STEM data acquired in vacuum.

In practice, noise in TEM images acquired using a charge coupled device (CCD) detector is mainly caused by the pixel-to-pixel sensitivity variation of the detector and is influenced by the electron dose. The effects of camera noise and electron dose have been explored on in Figure 5e. Since no contrast should be observed in vacuum, we assessed image quality by the standard deviation of phase images. As the beam current density increases, the influence of camera noise tends to decrease in reconstructions using both the CCD detector (before saturation) and the EMPAD detector. And the standard deviation of phase imaging using CCD detector in the presence of the electron beam is always larger than that in the absence of incident electrons (dark frames), suggesting that the noise in CCD detector mainly comes from the non-uniform sensitivity of the detector pixels. Meanwhile, the standard deviation in phase imaging using EMPAD detector is 4-5 orders of magnitude lower than that of using CCD detector. Although the SNR might be improved if TIE is performed using a direct detector, systematic misalignment from image registration cannot be avoided in the conventional method. Here, systematic misalignment includes image shifts and distortions in the optical path accompanying defocus changes (Figure Appendix B,S5) as well as random/personal/methodological errors during image registration. Note that an empirical low-pass filter can help improve the reconstruction quality using Lorentz-TEM images (see the inserted white box in Figure 5b) by compromising resolution but may introduce additional artifacts and errors.

## 4. Discussion

The BBD method is an efficient phase and field retrieval method in defocused STEM mode that utilizes the bright-field balanced divergency of the transmitted disk under the paraxial approximation.

The phases imaging using the BBD method is less dependent on atomic numbers and thus can be used to obtain atomic-resolution images of both light and heavy atoms. Our calculations show that the contrast is approximately proportional to $Z^{0.4}$. Compare to the phase imaging using image series acquired at different defocusing, the phase imaging using the BBD method are robust because the bright-field balanced divergency in the transmitted disk is independent of sample thickness and spherical aberrations. What's more, the BBD method is based on a single defocused transmitted disks and is therefore more dose-efficient than ADF-STEM that based on a limited angle of high-angle scattering angle ranges or methods based on defocused TEM image series. Meanwhile, a defocused probe used in BBD method can be compatible with phase reconstruction using mixed-state ptychographic approach for high-resolution and low-dose imaging [18]. Therefore, BBD method enable many potential applications from inorganic crystals to beam-sensitive organic materials.

For intermediate resolution phases, the projected magnetic/electric fields in the material can be imaged by BBD method using a small convergence angles. In experimental data, we compare PMP using the BBD method and conventional methods, including CoM-DPC method with 4D-STEM data and TIE method with defocused TEM image series. Due to the defocus dependence of PCTF of the virtual BF/ABF images used in BBD method, the PMP contrast can be distinguished from non-phase contrasts, which is approximately independent of defocuses. Thus, compared to the CoM-DPC method, the phase imaging using the BBD method is robust to transmitted disk shift or intensity changes associated with artificial structures or bending contours. Meanwhile, the defocus dependence of the PCTF of virtual BF/ABF images on the phase reconstruction means that weak phase signals can be enhanced by the defocusing setting. These features suggest that the BBD method is an alternative to the DPC method in phase and projected field retrieval, with better quality and effective information confinement (without background contrast blur).

The conventional TIE method requires acquiring a series of defocused TEM images. In practice, TEM images acquired at different defocusing and magnetic fields suffer from inevitable magnification changes, shifts, distortions, varying edge diffractions, and redistributed beam intensities due to changes in optical conditions. To perform phase retrieval using the TIE method, the acquired image series requires elaborate post-processing and analysis, including image registration and noise filtering that fundamentally introduce gross errors. In contrast, the BBD method is free of the aforementioned errors, as it is reconstructed from a single defocused 4D-STEM data, where the practical limitation is instrument and specimen stability, *i.e.* "scan noise". Furthermore, we compared the "camera noise phase" from TEM images acquired with a CCD detector and 4DSTEM acquired with an EMPAD detector. The standard deviation of the former is 4-5 orders of magnitude higher than the standard deviation of the latter used for the BBD method because of differences in pixel-to-pixel sensitivity in CCD detectors. 4D-STEM data for BBD methods also supports multipurpose reconstruction and analysis such as DPC, ptychography, virtual BF/ABF/ADF reconstruction and thickness/tilt mapping. It is noteworthy that the phase retrieval using the BBD method can also be conducted using segmented detectors designed for conventional BF and ABF imaging, without the need for complex iterative algorithms or empirical optimizations that consume a lot of computational power. With continuous improvements in instrument capacity and detector design, this technique can be integrated as an *in situ* automated imaging method, in parallel with BF/ABF/ADF and EELS imaging.

## 5. Conclusions

BBD method provides an efficient approach to get phase and field images from defocused STEM data. We have demonstrated the BBD method is very promising for improving the contrast of weakly scattering elements at atomic resolution, indicating potential applications for structural analysis in biological and physical sciences. Furthermore, these methods could be applied to the imaging of magnetic/electric phase with effectively high spatial resolution and robustness to non-phase background contrasts. In particular, the BBD method using a defocused electron probe is dose-efficient in terms of the intertwining intensity changes in the transmission disks at different scattering angles. When combined with fast direct electron detectors, the ability to image weakly scattering elements, high dose efficiency, noise-robustness, and compatibility with other STEM techniques make our method promising for exploring fundamental insights into many emerging materials candidates. For example, it can be explored for imaging weak phase transitions in 2D materials, and structural determination of heterogeneous soft/hard matter samples. The BBD method can also be applied to other scanning transfer techniques, such as X-ray and microscopy imaging systems, with potentially broad implications in related fields.

## Appendix A

### A1. Phase contrast transfer function (PCTF)

Consider a thin object of thickness t divided into M distinct slices of thickness $\Delta z$, where $\Delta z = t/M$. For an object slice at depth $\boldsymbol{z_i}$, the specimen transmission function (the complex amplitude distribution of the transmitted electron wave in the object plane) can be expressed as

$$T(\boldsymbol{r}, \boldsymbol{z_i}) = a(\boldsymbol{r}, \boldsymbol{z_i}) \exp[i\phi(\mathbf{r}, \mathbf{z_i})] = [a_0 + \Delta a(\boldsymbol{r}, \boldsymbol{z_i})] \exp[i\phi(\mathbf{r}, \mathbf{z_i})]$$

where **r** is the real space coordinates, $\phi(\boldsymbol{r}, \boldsymbol{z_i})$ is the phase of object in depth $\boldsymbol{z_i}$, $a_0$ represents the direct current component of the incident electron wave undisturbed by the object and $\Delta a(\boldsymbol{r}, \boldsymbol{z_i}) = a_0 \eta(\boldsymbol{r}, \boldsymbol{z_i})$ represents the contribution from the absorption variations. Using the weak phase approximation ($\phi(\mathbf{r}, \boldsymbol{z_i}) \ll 1$) and the weak absorption approximation ($\Delta a \ll a_0$), this can be simplified with coordinate conjugate of **k** in reciprocal space to:

$$T(\boldsymbol{k}, \boldsymbol{z_i}) \approx a_0[\delta(\boldsymbol{k}) + \eta(\boldsymbol{k}, \boldsymbol{z_i}) + i\phi(\mathbf{k}, \mathbf{z_i})].$$

For the incident electron wave $\Psi(\boldsymbol{k}, \boldsymbol{z_i}, \Delta\mathrm{f})$, the electron wave passing through the object slice can be expressed as the convolution of Fresnel free-space propagator $P(\boldsymbol{k}, \Delta z)$ and $[T(\boldsymbol{k}, \boldsymbol{z_i})\Psi(\boldsymbol{k}, \boldsymbol{z_i}, \Delta f)]$:

$$\Psi(\boldsymbol{k}, \boldsymbol{z_i} + \Delta z, \Delta\mathrm{f}) = P(\boldsymbol{k}, \Delta z) \otimes [T(\boldsymbol{k}, \boldsymbol{z_i})\Psi(\boldsymbol{k}, \boldsymbol{z_i}, \Delta\mathrm{f})],$$

where $P(\boldsymbol{k}, \Delta z) = \exp[-\pi i k^2 \lambda \Delta z]$ , and $\Psi(\boldsymbol{k}, \boldsymbol{z_i} = 0, \Delta\mathrm{f}) = \mathrm{A}(\boldsymbol{k}) \exp[-\mathrm{i}\chi(\boldsymbol{k}, \Delta\mathrm{f})]$. Here, the $\lambda$ is the electron wavelength, $\Delta\mathrm{f}$ is the defocus distance, $\mathrm{A}(\boldsymbol{k})$ is the pupil function of the objective aperture, and $\chi(\boldsymbol{k}, \Delta\mathrm{f})$ is the phase shift that has the form of

$$\chi(\boldsymbol{k}, \Delta\mathrm{f}) = \pi \Delta f \lambda |\boldsymbol{k}|^2 + \frac{\pi}{2} C_S \lambda^3 |\boldsymbol{k}|^4,$$

where $C_S$ is the spherical aberration. Thus, the image reconstructed from detector $D_\beta$ in STEM optical system can be expressed as

$$I_\beta(\boldsymbol{r}, \Delta\mathrm{f}, \Delta z, M) = \int |\Psi(\boldsymbol{k}, \boldsymbol{z_i} = \boldsymbol{t}, \Delta\mathrm{f})|^2 D_\beta(\boldsymbol{k}) d\boldsymbol{k} = I_0 + 2Re\left\{\sum_k \Xi(\boldsymbol{k}, \Delta\mathrm{f}, \Delta z, M) e^{2\pi i \boldsymbol{k}\cdot\boldsymbol{r}} \eta(\boldsymbol{k})\right\} + 2Re\left\{i \sum_k \Xi(\boldsymbol{k}, \Delta\mathrm{f}, \Delta z, M) e^{2\pi i \boldsymbol{k}\cdot\boldsymbol{r}} \phi(\boldsymbol{k})\right\},$$

where $\Xi(\boldsymbol{k})$ defined by

$$\Xi(\boldsymbol{k}, \Delta\mathrm{f}, \Delta z, M) = \sum_{m=0}^{M} [\Psi(\boldsymbol{k}, m\Delta z, \Delta\mathrm{f}) P(\boldsymbol{k}, m\Delta z)] \otimes [P^*(-\boldsymbol{k}, m\Delta z) \Psi^*(-\boldsymbol{k}, m\Delta z, \Delta\mathrm{f}) D_\beta(-\boldsymbol{k})],$$

and $I_0$ is defined by

$$I_0 = \sum_k |\Psi(\boldsymbol{k}, \boldsymbol{z_i} = 0, \Delta\mathrm{f})|^2 \mathrm{D}_\beta(\boldsymbol{k}).$$

Note the $\phi(\boldsymbol{k}, \boldsymbol{z_i})$ is assumed to be the same for different slices in the crystal samples for simplicity. Therefore, the integrated PCTF can be calculated as

$$\mathcal{L}_\beta(\boldsymbol{k}, \Delta f, \Delta z, M) = \frac{2}{I_0} Re\{i\Xi(\boldsymbol{k}, \Delta\mathrm{f}, \Delta z, M)\}.$$

### A2. Phase reconstruction using transport of intensity of equation (TIE) method in STEM mode

In the reciprocity theorem, BF/ABF images reconstructed in STEM optical system can correspond to a CTEM on reciprocity [29]. Therefore, phase reconstruction methods used for CTEM images, such as the TIE method, can also be used for BF/ABF images reconstructed in STEM mode. For clarity, we can simply express $I_\beta(\boldsymbol{r}, \Delta f)$ by setting M = 0, *i.e.* no slice. Then, we can have

$$\begin{aligned} I_\beta(\boldsymbol{r}, \Delta \mathrm{f}) &= I_0 + 2Re\{\textstyle\sum_k [\Psi(\boldsymbol{k}, \Delta \mathrm{f}) \otimes \Psi^*(-\boldsymbol{k}, \Delta \mathrm{f}) D_\beta(\boldsymbol{k})] e^{2\pi i \boldsymbol{k}\cdot\boldsymbol{r}} \eta(\boldsymbol{k})\} + \\ & 2Re\{i \textstyle\sum_k [\Psi(\boldsymbol{k}, \Delta \mathrm{f}) \otimes \Psi^*(-\boldsymbol{k}, \Delta \mathrm{f}) D_\beta(\boldsymbol{k})] e^{2\pi i \boldsymbol{k}\cdot\boldsymbol{r}} \phi(\boldsymbol{k})\} \\ &= I_0 - 2\{\textstyle\sum_k \cos[\chi(\boldsymbol{k}, \Delta \mathrm{f})] |A(\boldsymbol{k})|^2 D_\beta(\boldsymbol{k}) e^{2\pi i \boldsymbol{k}\cdot\boldsymbol{r}} \eta(\boldsymbol{k})\} \\ & -2\{\textstyle\sum_k \sin[\chi(\boldsymbol{k}, \Delta \mathrm{f})] |A(\boldsymbol{k})|^2 D_\beta(\boldsymbol{k}) e^{2\pi i \boldsymbol{k}\cdot\boldsymbol{r}} \phi(\boldsymbol{k})\} \end{aligned}$$

When $\frac{\pi}{2} C_S \lambda^3 |\boldsymbol{k}|^4 \ll \pi \Delta f \lambda |\boldsymbol{k}|^2$, *i.e.* the paraxial approximation $\lambda^2 |\boldsymbol{k}|^2 \ll 1$, the above expression can further be simplified as

$$\begin{aligned} I_\beta(\boldsymbol{r}, \Delta \mathrm{f}) \approx I_0 - 2\{\textstyle\sum_k \cos(\pi \Delta f \lambda |\boldsymbol{k}|^2) |A(\boldsymbol{k})|^2 D_\beta(\boldsymbol{k}) e^{2\pi i \boldsymbol{k}\cdot\boldsymbol{r}} \eta(\boldsymbol{k})\} - \\ 2\{\textstyle\sum_k \sin(\pi \Delta f \lambda |\boldsymbol{k}|^2) |A(\boldsymbol{k})|^2 D_\beta(\boldsymbol{k}) e^{2\pi i \boldsymbol{k}\cdot\boldsymbol{r}} \phi(\boldsymbol{k})\}. \end{aligned}$$

Therefore, to avoid the inverse PCTF, the defocus should be less than a critical number $\Delta f_c$ satisfying $\pi |\Delta f_c| \lambda |\boldsymbol{k}_{max}|^2 < \pi \Rightarrow |\Delta f_c| < \frac{1}{\lambda |\boldsymbol{k}_{max}|^2}$, where $\boldsymbol{k}_{max}$ is the information limit in reciprocal space. It is worth noting that if the defocus is reversed, the intensity changes due to the absorption term $\eta(\boldsymbol{k})$ is the same, but the phase component is opposite. Therefore, a pure phase contrast image can be calculated by subtracting the two images acquired at $\pm\Delta f$ [6]:

$$\phi(\boldsymbol{r}) = \frac{|I_\beta(\boldsymbol{r}, \Delta \mathrm{f}) - I_\beta(\boldsymbol{r}, -\Delta \mathrm{f})|}{4 I_\beta(\boldsymbol{r}, \Delta \mathrm{f} = 0) \sin(\pi \Delta f \lambda |\boldsymbol{k}|^2)} \xrightarrow{\Delta f \to 0} \frac{|I_\beta(\boldsymbol{r}, \Delta \mathrm{f}) - I_\beta(\boldsymbol{r}, -\Delta \mathrm{f})|}{4\pi \lambda \Delta f |\boldsymbol{k}|^2 [I_\beta(\boldsymbol{r}, \Delta \mathrm{f}) + I_\beta(\boldsymbol{r}, -\Delta \mathrm{f})]}$$

where $I_\beta(\boldsymbol{k}, \Delta f)$ is the Fourier transformed $I_\beta(\boldsymbol{r}, \Delta f)$. This is a form of TIE equation under the unify intensity since $\nabla^{-2} \leftrightarrow -1/4\pi^2 |\boldsymbol{k}|^2$ [6], *i.e.*

$$-\boldsymbol{k} \frac{\partial I(r)}{\partial \Delta f} = I(\boldsymbol{r}) \nabla^2 \phi(\boldsymbol{r}).$$

### A3. Bright-field balanced divergency (BBD) method

For a transmitted disk with convergence angle $\alpha$, we can set up two virtual detectors

$$\begin{cases} D_{\beta 1}(\boldsymbol{k}) = \begin{cases} 1, & \boldsymbol{k} < \beta 1 \\ 0, & \beta 1 < \boldsymbol{k} < \alpha \end{cases} \\ D_{\beta 2}(\boldsymbol{k}) = \begin{cases} 0, & \boldsymbol{k} < \beta 1 \\ 1, & \beta 1 < \boldsymbol{k} < \alpha \end{cases} \end{cases}$$

For simplicity, images reconstructed from these two detectors of single slice object under the paraxial approximation can be wrote as

$$\begin{cases} I_{\beta 1}(\boldsymbol{r}, \Delta f) = I_{0,\beta_1}(\boldsymbol{k}) - 2\left\{\int_0^{\beta_1} \cos(\pi \Delta f \lambda |\boldsymbol{k}|^2) e^{2\pi i \boldsymbol{k}\cdot\boldsymbol{r}} \eta(\boldsymbol{k}) d\boldsymbol{k}\right\} - 2\left\{\int_0^{\beta_1} \sin(\pi \Delta f \lambda |\boldsymbol{k}|^2) e^{2\pi i \boldsymbol{k}\cdot\boldsymbol{r}} \phi(\boldsymbol{k}) d\boldsymbol{k}\right\} \\ I_{\beta 2}(\boldsymbol{r}, \Delta f) = I_{0,\beta_2}(\boldsymbol{k}) - 2\left\{\int_{\beta_1}^{\alpha} \cos(\pi \Delta f \lambda |\boldsymbol{k}|^2) e^{2\pi i \boldsymbol{k}\cdot\boldsymbol{r}} \eta(\boldsymbol{k}) d\boldsymbol{k}\right\} - 2\left\{\int_{\beta_1}^{\alpha} \sin(\pi \Delta f \lambda |\boldsymbol{k}|^2) e^{2\pi i \boldsymbol{k}\cdot\boldsymbol{r}} \phi(\boldsymbol{k}) d\boldsymbol{k}\right\} \end{cases}$$

Here, $\cos(\pi \Delta f \lambda |\boldsymbol{k}|^2)$ can be expanded in its Taylor series format:

$$\cos(\pi \Delta f \lambda |\boldsymbol{k}|^2) = 1 - \frac{(\pi \Delta f \lambda |\boldsymbol{k}|^2)^2}{2!} + \frac{(\pi \Delta f \lambda |\boldsymbol{k}|^2)^4}{4!} - \cdots$$

When $\Delta f$ or **k** is small, we can ignore higher-order expansion terms in the paraxial approximation $\lambda^2|\boldsymbol{k}|^2 \ll 1$, *i.e.* $\cos(\pi\Delta f\lambda|\boldsymbol{k}|^2) \approx 1$. Thus, the expression for the reconstructed image from these two detectors can be simplified to

$$\begin{cases} I_{\beta 1}(\boldsymbol{r},\Delta f) = I_{0,\beta_1}(\boldsymbol{k}) - 2\left\{\int_0^{\beta_1} e^{2\pi i\boldsymbol{k}\cdot\boldsymbol{r}}\eta(\boldsymbol{k})d\boldsymbol{k}\right\} - 2\left\{\int_0^{\beta_1}\sin(\pi\Delta f\lambda|\boldsymbol{k}|^2)e^{2\pi i\boldsymbol{k}\cdot\boldsymbol{r}}\phi(\boldsymbol{k})d\boldsymbol{k}\right\} \\ I_{\beta 2}(\boldsymbol{r},\Delta f) = I_{0,\beta_2}(\boldsymbol{k}) - 2\left\{\int_{\beta_1}^{\alpha} e^{2\pi i\boldsymbol{k}\cdot\boldsymbol{r}}\eta(\boldsymbol{k})d\boldsymbol{k}\right\} - 2\left\{\int_{\beta_1}^{\alpha}\sin(\pi\Delta f\lambda|\boldsymbol{k}|^2)e^{2\pi i\boldsymbol{k}\cdot\boldsymbol{r}}\phi(\boldsymbol{k})d\boldsymbol{k}\right\} \end{cases}$$

and when $\Delta f = 0$, we have

$$\begin{cases} I_{\beta 1}(\boldsymbol{r},\Delta f = 0) = I_{0,\beta_1}(\boldsymbol{k}) - 2\left\{\int_0^{\beta_1} e^{2\pi i\boldsymbol{k}\cdot\boldsymbol{r}}\eta(\boldsymbol{k})d\boldsymbol{k}\right\} \\ I_{\beta 2}(\boldsymbol{r},\Delta f = 0) = I_{0,\beta_2}(\boldsymbol{k}) - 2\left\{\int_{\beta_1}^{\alpha} e^{2\pi i\boldsymbol{k}\cdot\boldsymbol{r}}\eta(\boldsymbol{k})d\boldsymbol{k}\right\} \end{cases}$$

Since the phase part only determines the contrast distribution in the Ronchigram, the above equation show the total intensity, $I_{sum}$, of these two detectors is approximately constant at each scan position, dependent on $\eta(\boldsymbol{k})$ but independent of $\Delta f$, *i.e.*

$$I_{sum} = I_{\beta 1}(\boldsymbol{r},\Delta f) + I_{\beta 2}(\boldsymbol{r},\Delta f) = I_{\beta 1}(\boldsymbol{r},\Delta f = 0) + I_{\beta 2}(\boldsymbol{r},\Delta f = 0).$$

Therefore, the phase contrast signal in $D_{\beta 1}(\boldsymbol{k})$ and $D_{\beta 2}(\boldsymbol{k})$ is “bright-field balanced”,

$$\left\{\int_0^{\beta_1}\sin(\pi\Delta f\lambda|\boldsymbol{k}|^2)e^{2\pi i\boldsymbol{k}\cdot\boldsymbol{r}}\phi(\boldsymbol{k})d\boldsymbol{k}\right\} + \left\{\int_{\beta 1}^{\alpha}\sin(\pi\Delta f\lambda|\boldsymbol{k}|^2)e^{2\pi i\boldsymbol{k}\cdot\boldsymbol{r}}\phi(\boldsymbol{k})d\boldsymbol{k}\right\} = 0$$

$$\Rightarrow \left\{\int_0^{\beta_1}\sin(\pi\Delta f\lambda|\boldsymbol{k}|^2)e^{2\pi i\boldsymbol{k}\cdot\boldsymbol{r}}\phi(\boldsymbol{k})d\boldsymbol{k}\right\} = \left\{\int_{\beta 1}^{\alpha}\sin(\pi(-\Delta f)\lambda|\boldsymbol{k}|^2)e^{2\pi i\boldsymbol{k}\cdot\boldsymbol{r}}\phi(\boldsymbol{k})d\boldsymbol{k}\right\}$$

Compare with phase reconstruction using TIE method with defocused image series, we can calculate the phase by $I_{\beta 1}(\boldsymbol{k},\Delta f)$ and $I_{\beta 2}(\boldsymbol{k},\Delta f)$ when $\Delta f$ is small using

$$\phi(\boldsymbol{r}) \approx \frac{|\gamma * I_{\beta 1}(\boldsymbol{r},\Delta f) - I_{\beta 2}(\boldsymbol{r},\Delta f)|}{2*(1+\gamma)[I_{\beta 1}(\boldsymbol{r},\Delta f) + I_{\beta 2}(\boldsymbol{r},\Delta f)]\sin(\pi\Delta f\lambda|\boldsymbol{k}|^2)}$$

$$= \frac{|\gamma * I_{\beta 2}(\boldsymbol{r},-\Delta f) - I_{\beta 2}(\boldsymbol{r},\Delta f)|}{2*(1+\gamma)[I_{\beta 2}(\boldsymbol{r},-\Delta f) + I_{\beta 2}(\boldsymbol{r},\Delta f)]\sin(\pi\Delta f\lambda|\boldsymbol{k}|^2)}$$

where $\gamma = \frac{I_{\beta 2}(\boldsymbol{r},\Delta f=0)}{I_{\beta 1}(\boldsymbol{r},\Delta f=0)} \approx \frac{I_{\beta 2}(\boldsymbol{r},\Delta f)}{I_{\beta 1}(\boldsymbol{r},\Delta f)}$ in weak phase approximation. Further, the gradients of magnetic flux density $B_{(y,x)}(\boldsymbol{r})$ can be calculated from the gradients of magnetic phase using

$B_{(y,x)}(\boldsymbol{r}) = -\frac{h}{2\pi e}\frac{\partial\phi(r)}{\partial(x,y)}\, t^{-1}(\boldsymbol{r})$ , where $t(\boldsymbol{r})$ is the sample thickness.

**Acknowledgments:** We thank Dr. Núria Bagués for providing the geometric confined FeGe sample and helpful discussion. **Fundings**: This work was primarily supported by DARPA under Grant No. D18AP00008. Binbin Wang also thanks support from Presidential Fellowship of the Ohio State University. DWM acknowledges partial support from the Center of Emergent Materials, an NSF MRSEC, under award number DMR-2011876. Electron microscopy experiments were supported by the Center for Electron Microscopy and Analysis at the Ohio State University. MuSTEM simulations were partially supported by an allocation of computing time from the Ohio Supercomputer Center. **Author Contributions:** B.W. and D.M. designed research; B.W. performed research; B.W. analyzed data; and B.W. and D.M. wrote the paper. Competing interests: Authors declare no competing interests; and **Data and materials availability**: Scripts and data can be found in https://github.com/wang10255/Bright-field-Balanced-Method.